\documentclass[twocolumn,superscriptaddress,secnumarabic,amsmath,amssymb, nobibnotes, aps, prd]{revtex4-1}
\usepackage{graphicx}
\usepackage[english]{babel}
\usepackage[utf8x]{inputenc}
\usepackage{dcolumn}
\usepackage[colorinlistoftodos]{todonotes}
\usepackage{xcolor}
\usepackage{mathtools}
\usepackage{bm}

\usepackage{hyperref}

\begin{document}

\preprint{APS/123-QED}


\title{Interacting color strings as the origin of the liquid behavior of the quark-gluon plasma}

\author{J. E. Ramírez}
\email{jhony.ramirezcancino@viep.com.mx}
\affiliation{
Departamento de F\'isica de Part\'iculas and Instituto Galego de Física de Altas Enerxías, Universidad de Santiago de Compostela, E-15782 Santiago de Compostela, Espa\~na}
\affiliation{Centro de Agroecología,
Instituto de Ciencias,
Benemérita Universidad Autónoma de Puebla, Apartado Postal 165, 72000 Puebla, Puebla, M\'exico}
\author{Bogar Díaz}
\email{bodiazj@math.uc3m.es}

\affiliation{Departamento de Matem\'aticas, Universidad Carlos III de Madrid. Avenida  de la Universidad 30, 28911 Legan\'es, Spain}
\affiliation{Departamento de Física de Altas Energías, Instituto de Ciencias Nucleares, Universidad Nacional Autónoma de México, Apartado Postal 70-543, Ciudad de México, 04510, México} 
\affiliation{Grupo de Teorías de Campos y Física Estadística. Instituto Gregorio Millán (UC3M), Unidad Asociada al Instituto de Estructura de la Materia, CSIC, Serrano 123, 28006 Madrid, Spain}
\author{C. Pajares}
\email{pajares@fpaxp1.usc.es}
\affiliation{
Departamento de F\'isica de Part\'iculas and Instituto Galego de Física de Altas Enerxías, Universidad de Santiago de Compostela, E-15782 Santiago de Compostela, Espa\~na}


\begin{abstract}


We study the radial distribution function of the color sources (strings) formed in hadronic collisions and the requirements to obtain a liquid. As a repulsive interaction is needed, we incorporate a concentric core in the strings as well as the probability that a string allows core-core overlaps. We find systems where the difference between the gas-liquid and confined-deconfined phase transition temperatures is small. This explains the experimentally observed liquid behavior of the quark-gluon plasma above the confined-deconfined transition temperature.
\end{abstract}

\maketitle


\section{Introduction}
The heavy-ion Au-Au collisions at RHIC obtained a liquid of quarks and gluons with a shear viscosity over entropy density ratio lower than any other material ever known (quark-gluon plasma) \cite{gyulassy, ADAMS2005102,ADCOX2005184}. This discovery was confirmed later in Pb-Pb collisions at LHC through the study of all of the harmonics of the azimuthal distributions and different correlations showing the existence of a collective motion of quarks and gluons \cite{PhysRevLett.105.252302,2012330,CMS}. Most of the properties seen in heavy-ion collisions have also been observed in pp, pA collisions at LHC \cite{CMS2}, and dAu and $^3$HeAu at RHIC \cite{Aidala2019}.

Multiparticle production in pp, pA, and AA collisions is currently described in terms of color strings stretched between the partons of the projectile and target, which decay into new strings and subsequently into hadrons. The strings are extended in the longitudinal space between the partons of the colliding projectiles, which transforms it into a rapidity space describing the rapidity differences of the partons. This difference is given by the energy fraction of the projectiles carried by the partons. On the other hand, the strings also have a transverse extension. Thus, color strings may be viewed as small areas in the transverse plane of the collision filled with color field created by the colliding partons. Due to confinement, the strings have transverse circular areas with radius $r_0$=$\sigma/2$ around 0.2-0.3 fm ($\sigma$ is the diameter). With the growing energy or size of the colliding systems, the number of strings grows, and they start to overlap and interact. In the color string percolation model (CSPM) the interaction between strings occurs forming clusters when they overlap; the color field is given by the SU(3) color sum. Due to the random direction of the color field in the color space, the intensity of the resulting color field is only the square root of the individual strings color field \cite{BRAUN20151,string,Braun2000,PRLPajares}. In rapidity, the interaction of strings is taken into account by incorporating the energy-momentum conservation of the formed clusters. This gives rise to the particle production in the forward region outside the kinematical limits, the so-called cumulative effect, observed at RHIC energies \cite{ARMESTO199678, Braun20022,ANDYC}. However, in our study, we are only concerned with the  transverse size.

At one critical string density a spanning cluster is formed through the collision surface. For a uniform profile of the string distribution, this critical density value, in the thermodynamic limit, matches the percolation threshold of the classical two dimensional (2D) continuum percolation model, given by $N/S$=1.128/$\pi r_0^2$, where $N$ and $S$ are the number of strings and the collision area respectively. 
This critical value can change slightly for a finite and not large $N$ and other profiles \cite{RODRIGUES1999402,RAMIREZ20178, jerc}. This critical percolation density is associated with a temperature $T= 160$ MeV, which corresponds to the confined-deconfined phase transition of the quark and gluon matter \cite{DIASDEDEUS2006455}. 

\begin{figure}
\centering
\includegraphics[scale=1]{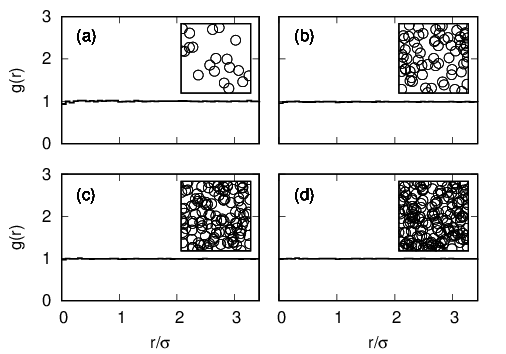}
\caption{Radial distribution function for the CSPM at different density values ($\pi r_0^2 N/S$): (a) 0.32, (b) 1.12, (c) 1.92, and (d) 2.72.}
\label{fig:cspm-gr}
\end{figure}

On the other hand, the physical structure of the systems can be determined by analyzing the behavior of the radial distribution function, $g(r)$ (also known as the pair correlation function). 
It describes how the average number of particles varies as a function of the radial distance from a point.
It is widely used to characterize packing structures and contains information about long-range interparticle correlations and their organization \cite{ASTE200632}.

Since models considering strings like fully penetrable disks correspond to the picture of the classical ideal gas, it is expected that they will have a flat radial distribution function. In Fig.~\ref{fig:cspm-gr}, we show for the CSPM the flat behavior of $g(r)$, regardless of the string density. This happens even when the CSPM can explain most of the experimental data on pp, pA, and AA collisions, including the azimuthal distributions of the produced particles,  
as well as the temperature dependence of the ratio between the shear and bulk viscosities over the entropy density \cite{P2016, Sahoothermo}.
This ideal gas behavior of the CSPM has been observed by some authors. For example, in Refs.~\cite{sahoo, sahoo2} the authors analyzed the electrical and thermal conductivity of the quark-gluon plasma (QGP) using the CSPM and found that the system corresponds to an ideal gas. Also, in Ref.~\cite{jerc} the authors performed a finite-size analysis of the speed of sound and concluded that the CSPM corresponds to a mean field theory. On the other hand,
there are other strings models \cite{ridge, pythia,Bierlich2015} where the strings interact in a different way, for instance, by color rearrangements \cite{pythia} or by shoving when they are close to each other \cite{Bierlich2015}. 
The repulsion between strings has also been used to study the harmonics of the azimuthal distributions, obtaining a reasonable agreement with the data
\cite{Kalayd, aip}.

In this paper, we propose a hybrid core-shell model together with the traditional CSPM to incorporate the excluding or repulsive interaction between strings. We do this by providing the strings with a concentric region of exclusion (core region) of diameter $\lambda \sigma$ ($0\leq\lambda\leq 1$). The rest of the string area is called the shell region. We also consider a probability $q_\lambda$ that a string allows overlap in its core with the core of another string. We refer to the strings as soft or hard if they admit overlap in their core region or not, respectively. 
Notice that the overlap condition applies only to the core-core interactions, while all core-shell and shell-shell overlaps are allowed. In what follows, we call this modification the core-shell-color string percolation model (CSCSPM). Notice that i) if $(\lambda=1, q_\lambda=0)$ the system corresponds to a hard-disks fluid \cite{chandler}, ii) if $\lambda=0$ or $q_\lambda=1$ the system returns to the traditional 2D continuum percolation \cite{continuumperc,mertens}, iii) if $q_\lambda=0$ this model reproduces the continuum percolation of disks with hard cores \cite{hcperc,hcperc2}, and iv) if $\lambda=1$ it resembles the random sequential absorption model \cite{rejection,rejection2}. In this sense, our model is a generalization of them.
For the CSCSPM, two phase transitions are observed as a function of $(\lambda, q_\lambda)$: ideal gas to non-ideal to liquid.
This paper aims to determine if some combinations of $(\lambda, q_\lambda)$ allow the system to simultaneously exhibit the gas-liquid transition and the emergence of the spanning cluster. To this end, we determine: a) the gas-liquid phase transition temperature, which is calculated through the observation of the oscillation of the radial distribution function, and b) the critical temperature associated with the percolation threshold, which corresponds to the QGP formation temperature.

The plan of the paper is as follows. In Sec.~\ref{sec:temp}, we present how to calculate the temperature in percolation-based string models. In Sec. ~\ref{sec:simulation}, we provide the simulation and data analysis methods. In Sec.~\ref{sec:results}, we show the phase diagram of the CSCSPM in terms of $(\lambda, q_\lambda$), and  determine the region in the $\lambda$-$q_\lambda$ plane where the transition and critical temperatures are close.
Finally, Sec.~\ref{sec:conclusions} contains our conclusions and perspectives.


\section{Temperature for percolation-based string models}\label{sec:temp}

The interaction among strings gives rise to a reduction in multiplicity and an increase in the average transverse momentum.
A cluster of $n$ strings (remember that each string is considered as a disk) that occupies an area $S_n$ behaves as a single color source with a higher color field $\vec{Q}_n$ corresponding to the vectorial sum of the color charges of each individual string $\vec{Q}_1$. As $\vec{Q}_n=n\vec{Q}_1$ and the individual string colors may be oriented in an arbitrary manner respective to each other, the average $\vec{Q}_{1i}\vec{Q}_{1j}$ is zero and $\vec{Q}_n^2=n\vec{Q}_1^2$. Knowing the color charge, we can calculate the multiplicity $\mu_n$ and the mean transverse momentum squared $\langle p_T^2\rangle_n$ of the particles produced by a cluster, which are proportional to the color charge and color field respectively, as
\begin{subequations}
\begin{align}
\mu_n & =  \sqrt{\frac{nS_n}{S_1}}\mu_1 \, ,\label{multiplicity}\\
\langle p_T^2 \rangle_n & =  \sqrt{\frac{nS_1}{S_n}}\langle p_T^2 \rangle_1\, ,\label{momentum}
\end{align}
\end{subequations}
where $\mu_1$ and $\langle p_T^2 \rangle_1$ are the multiplicity and transverse momentum squared of the particles produced from a single string with transverse area $S_1=\pi r_0^2$.
For strings just touching each other, $S_n=nS_1$ and therefore $\mu_n=n\mu_1$ and $\langle p_T^2 \rangle_n=\langle p_T^2 \rangle_1$.
On the other hand, when strings fully overlap, $S_n=S_1$ and therefore $\mu_n=\sqrt{n}\mu_1$ and $\langle p_T^2 \rangle_n=\sqrt{n}\langle p_T^2 \rangle_1$, so that the multiplicity is maximally suppressed and the transverse momentum is maximally enhanced.
In the thermodynamic limit, the average value of $nS_1/S_n$ for all of the clusters is \cite{Braun2000}
\begin{equation}
    \left\langle \frac{nS_1}{S_n} \right\rangle =\frac{\eta}{1-e^{-\eta}} =\frac{1}{F(\eta)^2} \,,
\end{equation}
with $\eta=NS_1/S$ being the filling factor, where $N$ and $S$ are the number of strings and total surface area, respectively.
The function $F(\eta)$ is called the color reduction factor, which can be written in terms of the area covered by strings $\phi(\eta)$ as
\begin{equation}
    F(\eta)=\sqrt{\frac{\phi(\eta)}{\eta}}\,.
\end{equation}
In the thermodynamic limit, for fully penetrable disks \cite{BRAUN20151, Kertesz1982},
\begin{equation}
    \phi(\eta)=1-e^{-\eta}\,.
\end{equation}
However, the explicit form of $\phi$ depends on the considered string model, e.g., if the system is finite or without periodic boundary conditions with a particular geometry \cite{jerc} or, as in our case, if the system is composed of a combination of disks allowing overlap or not in the core region.
Now, Eqs.~\eqref{multiplicity} and \eqref{momentum} can be written as
\begin{subequations}
\begin{align}
\mu&=NF(\eta)\mu_1 \,,\\
\langle p_T^2 \rangle&= \frac{\langle p_T^2 \rangle_1}{F(\eta)}\,.
\end{align}
\end{subequations}

We must recall that the strings decay into new ones through color neutral $q-\bar{q}$ or $qq-\bar{q}\bar{q}$ pair production. The Schwinger QED$_2$ string-breaking mechanism produces these pairs at the time $\tau\sim1$fm/c, which subsequently hadronize to produce the observed hadrons.
The Schwinger distribution for the produced particles is $dN/dP_T^2\sim \exp(-p_T^2\pi/x²)$, where the average value of the string tension (color field) is $\langle x^2 \rangle$. The chromoelectric field in the string is not constant but fluctuates around its average value. Such fluctuations lead to a Gaussian distribution for the chromoelectric field, transforming the Schwinger distribution into the thermal distribution \cite{DIASDEDEUS2006455}
\begin{equation}
    \frac{dN}{dp_T^2}\sim \exp \left( -p_T \sqrt{\frac{2\pi}{\langle x^2 \rangle}}  \right)\,, \label{eq:thermal}
\end{equation}
with $\langle x^2 \rangle=\pi \langle p_T^2 \rangle_1/F(\eta)$.
Equation \eqref{eq:thermal} shows that the effective temperature is
\begin{equation}
    T(\eta)=\sqrt{\frac{\langle p_T^2 \rangle_1}{2F(\eta)}}\label{eq:temperatura}\,.
\end{equation}
This temperature is related to the Hawking-Unruh effect \cite{hawking, Unruh}.
In AA and pp collisions the thermalization can occur through the existence of an event horizon due to a rapid deceleration of the colliding nuclei \cite{Castorina2007}. Barrier penetration of the event horizon leads to a partial loss of information and is the reason for the stochastic thermalization of the $q-\bar{q}$ pairs.
In string percolation, as clusters grow and cover most of the collision area, this local temperature can be regarded as the temperature of the total thermal distribution.
In the 2D continuum percolation, for a uniform density profile, the critical filling factor at which the spanning cluster emerges is $\eta_c\sim$1.128 \cite{mertens}. This value can change slightly for a finite $N$ and other profiles \cite{RODRIGUES1999402,RAMIREZ20178, jerc}.
Introducing this value into Eq.~\eqref{eq:temperatura}, for a reasonable value of $\sqrt{\langle p_T^2 \rangle_1}$, around 200 MeV, we obtain a critical temperature $T_c$ around 160 MeV, which corresponds to the confined-deconfined phase transition of the quark and gluon matter \cite{DIASDEDEUS2006455}.


\begin{figure*}
\centering
\includegraphics[scale=1]{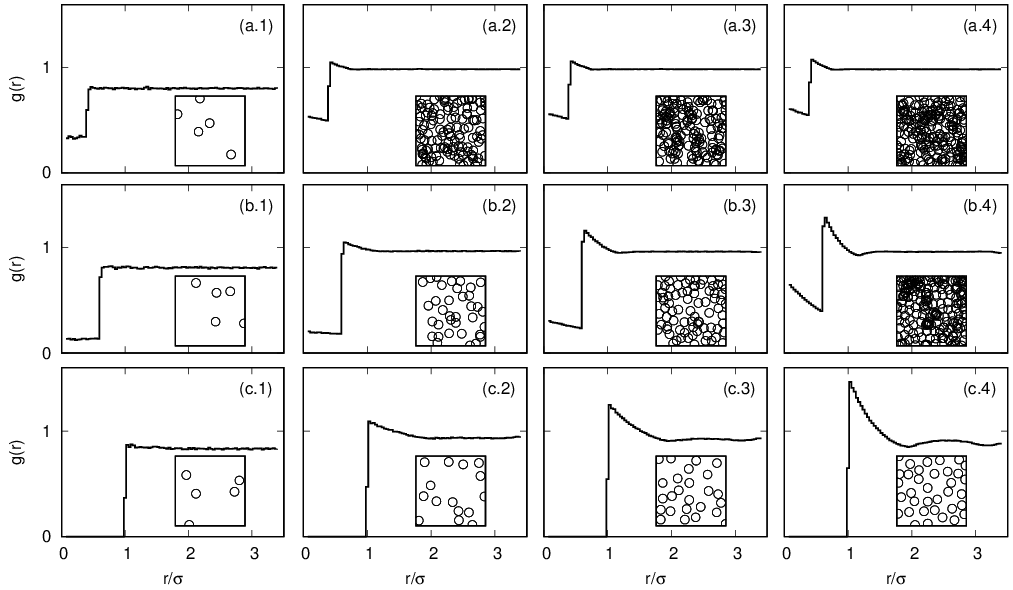}
\caption{Examples of the structures of core-shell-color string systems. We show $g(r)$ for three different pairs $(\lambda, q_\lambda)$ [(a) ($\lambda=0.4, q_\lambda=0.65$), (b)  ($\lambda=0.6, q_\lambda=0.4$), and (c)  ($\lambda=1, q_\lambda=0$)] and how it changes as $\eta$ varies. In panels (a.1), (b.1), and (c.1), the filling factor is $\eta=0.061$. These systems are diluted and show an ideal gas structure. In panels (a.2), (b.2), and (c.2), the filling factor corresponds to the transition from an ideal gas to a non-ideal gas ($\eta=$1.534, 0.429, and 0.184, respectively). In panels (b.3) and (c.3) the filling factor is $\eta=$0.982 and 0.307, respectively. These cases are when we observe the transition from non-ideal gas to a liquid-like structure. In panels (a.3) and (a.4) ($\eta=$ 0.981 and 2.454, respectively), the systems still show the non-ideal gas structure.. Finally, in panels (b.4) and (c.4) we have the maximum possible value of $\eta$  (2.454 and 0.429,  respectively) and these systems possess a liquid-like structure.
}
\label{fig:ejemplos}
\end{figure*}

\section{Simulation method and data analysis}\label{sec:simulation}

In this section, we discuss the computational implementation to calculate the radial distribution function $g(r)$ which is the average number density at a radial distance from a tagged string relative to the ideal gas case ($N/L^2$).
The plot of this function reveals the phase at which the system is \cite{chandler}. 
In the case of an ideal gas, it corresponds to a flat function, indicating that the system is well distributed and that any disk may be at any place. 
Another situation corresponds to a diluted system composed of interacting disks.  
In this case, $g(r)$ shows a global maximum around the distance $\lambda \sigma$ as a consequence of the (short-distance) repulsive interaction. 
Nevertheless, as $r$ increases, $g(r)$ becomes flat due to the low number density. 
Finally, if the number density increases, the distance between the strings becomes smaller, the particle interactions are more frequent, and the radial distribution function starts to oscillate. 
This behavior occurs because the particles settle around each other. From the perspective of a tagged particle, it looks to be surrounded by a first shell of particles at a distance equivalent to the repulsive interaction range. 
They are the nearest neighbors. 
Subsequently, if the number density is high enough, a second shell could take arise: the next-to-nearest neighbors. 
Thus, particle interactions give rise to a structured system, identified as a liquid (observed in x-ray scattering \cite{xraysca} and non-vibrating magnetic granular systems \cite{donado} experiments).

In the computational implementation, we use the random sequential addition algorithm \cite{rsaddition} to add disk by disk, with the corresponding $(\lambda, q_\lambda)$. The simulation process is stopped according to the observable under calculation.

To determine $g(r)$ we add until 200 strings (if possible) distributed into a square box of size $L=8\sigma$ and randomly assigned as soft or hard according to $q_\lambda$. The first  added string is allocated in the geometrical center of the box and it is taken as a trial disk. Then, $g(r)$ is calculated as
\begin{equation}
    g(r)=\frac{n(r)L^2}{N(2\pi \Delta r (r+0.5\Delta r)+\pi\Delta r^2)}\,,
\end{equation}
where  $n(r)$ is the average number of strings at a distance between $r$ and $r+\Delta r$ from the trial disk \cite{hcperc, hcperc2}. It is computed over $1\times10^6$ simulated systems, starting with $N=5$ and increasing it in steps of $\Delta N=5$, and $\Delta r=0.035$.


We classify the pair values $(\lambda,q_\lambda)$ for each $\eta$ according to $g(r)$ and its derivative with respect to $r$, $g'(r)$, which is calculated using the five-point stencil method (with a spacing between points of $\Delta r$). The classification criteria are as follows (cf. Ref. \cite{chandler}):
\begin{enumerate} \label{clastruc}
    \item Ideal gas: if $g(r)$ is a flat function.
    \item Non-ideal gas: if $g(r)$ has a global maximum around $r_1= \lambda \sigma$ and $g'(r)$ has a sustained negative trend for $\lambda<r/\sigma<2\lambda$.
    \item Liquid-like structure: if the system has already shown the non-ideal gas structure for some $\eta$, and if for a higher density, $g(r)$ has a local minimum before  $2\lambda$, and  $g'(r)$ has a sustained positive trend for $2\lambda< r/\sigma < 2.5 \lambda$.
\end{enumerate}
The general characteristic contained in criteria 1, 2, and 3 are based on the behavior of $g(r)$ for classical fluids \cite{chandler}. In particular, the details of  the third point have been set by observing the oscillation behavior of $g(r)$ for the hard-sphere case $(\lambda=1, q_\lambda=0)$. This is relevant since it corresponds directly with the classical fluid of hard spheres and the oscillations of the radial distribution functions are the indication that the liquid-like structure has taken place \cite{chandler}.
The first maximum is originated by the next nearest neighbors around the tagged particle, the first coordination shell, which for the case of a hard string is at least to a radial distance $\lambda \sigma$.
Meanwhile, the second maximum corresponds to the location of the next-to-nearest neighbors. 
We do this because for other values of ($\lambda, q_{\lambda}$)  we do not expect to exactly reproduce the radial distribution function for a classical fluid since, in this work, $g(r)$ is computed as the average over system simulations where the tagged string may be soft or hard. 
In Fig.~\ref{fig:ejemplos} we show some examples of this classification  for different pairs $(\lambda, q_\lambda)$.

We analyze $g(r)$ in a segment of length $\lambda\sigma$, but as $r$ takes discrete values, it is necessary to establish a condition on how many consecutive points we are going to check. We select five points; this implies that  a transition from an ideal gas to a non-ideal gas is detected if  $\lambda \geq  0.2$. The classification begins with the pair $(\lambda=1,q_\lambda=0)$ (hard disk fluid limit) and we search if the system shows a liquid-like structure for some $\eta$.
Taking into account that the systems should approach to the ideal gas case as $\lambda$ decreases and $q_{\lambda}$ increases, we determine that if for some $(\lambda_0,q_{\lambda 0})$ the system does not manifest the liquid-like structure for all $\eta$, then for $(\lambda,q_{\lambda 0})$ with $\lambda<\lambda_0$ or $(\lambda_0,q_{\lambda})$ with $q_\lambda> q_{\lambda 0}$ it will not show the liquid-like structure anymore.
We use the same considerations for the transition from a non-ideal gas to an ideal gas. 


We use the Mertens-Moore method \cite{mertens} to determine the percolation threshold in the CSCSPM. Unlike the determination of $g(r)$, now we add disks until the emergence of the spanning cluster and save the number $n_c$ of added disks. Then, using the information of  $10^4$ simulated systems, we calculate the probability, $f_L(n)$, that a spanning cluster exists when $n$ disks have been added. The percolation probability $P_L(\eta)$ at $\eta$ is computed (as in Ref.~\cite{mertens}) by convoluting the $f_L(n)$ probability with the Poisson distribution with average $\alpha=\eta L^2/\pi r_0^2$, for several $\eta$ values around the maximum of the distribution of $\eta_c$. Then, each  data set is fitted to the function 
\begin{equation}
P_L(\eta)=\frac{1}{2}\left(1+\tanh \left( \frac{\eta-\eta_{cL}}{\Delta_L}  \right)   \right),
\end{equation}
where $\eta_{cL}$ is the estimated percolation threshold of the system of size $L$, and $\Delta_L$ is the amplitude of the transition region \cite{SABERI20151, Rintoul_1997}.
To take into account the finite-size effects on $\eta_{cL}$, we perform simulations with systems of size $L$=12, 16, 24, 32, 48, 64, and 96. Furthermore, we determine the percolation threshold in the thermodynamic limit, $\eta_c$, through the finite-size scaling $\eta_c-\eta_{cL}\propto \Delta_L^{-1/\nu}$. 
Here, $\nu$ is the exponent associated with the scaling of the correlation length of the cluster size, which is calculated using $\Delta_L\propto L^{-1/\nu}$ \cite{Rintoul_1997, stauffer}. From the analysis of $\Delta_L$ (as a function of $L$), we find $\nu\sim 4/3$ for all considered pairs ($\lambda, q_\lambda$). This is in good agreement with the results of 2D percolating systems \cite{stauffer}.


To measure the area covered by strings  we draw a square grid with spacing $L/20$. Then, we count the cells whose center is inside of at least one disk. In this way, the area is approximated as $A_L=\mathcal{N}L^2/400$, where $\mathcal{N}$ is the number of counted cells. 
We measure the area in the conditions: a) the density of the non-ideal gas to liquid-like transition, and b) the emergence of the spanning cluster. In particular, for situation b), it is possible to compute the area in the thermodynamic limit. In this case, we determine the mean area $A_{L}(n_c)$ that covers the $n_c$ strings. 
Then, the area covered by strings in the percolation threshold, $\phi(\eta_c)$, is calculated as the convolution of $A_{L}(n_c)$ with the Poisson distribution with average $\alpha_c=\eta_{cL}L^2/\pi r_0^2$. The finite-size effects on $\phi_L(\eta_c)$ satisfy  $\phi(\eta_c)-\phi_L(\eta_{cL})\propto L^{-1/\nu}$ [this relation is no longer valid as $(\lambda, q_\lambda)\rightarrow (1,0)$, this zone does not belong to the cases discussed above].


\section{Results}\label{sec:results} 
To analyze the deviation between the gas-liquid transition temperature [$T_c^*:=T(\eta_c^*)$] and the critical transition temperature [$T_c:=T(\eta_c)$] for a  given $(\lambda, q_\lambda)$, we define
\begin{equation}
\tau:=\left| \frac{T_c-T_c^*}{T_c}  \right|=\left| 1-\left( \frac{\eta_c^*\phi(\eta_c)}{\eta_c \phi(\eta_c^*)} \right)^{1/4}  \right| \,,
\label{eq:Tnorm}
\end{equation}
where $\eta_c^*$ is the minimum density at which the system shows a liquid-like structure, while $\eta_c$ is the percolation threshold.
Notice that: a) $\tau$ is independent of $\langle p_T^2 \rangle_1$ [which may dependent on ($\lambda$, $q_\lambda$) and on the size of the system], b) it only depends on the critical filling factors, $\eta_c$ and $\eta_c^*$, and the corresponding covered area, and c) it can be interpreted as the relative deviation between $T_c^*$ and $T_c$. 

\begin{figure}[ht]
\centering
\includegraphics[scale=1]{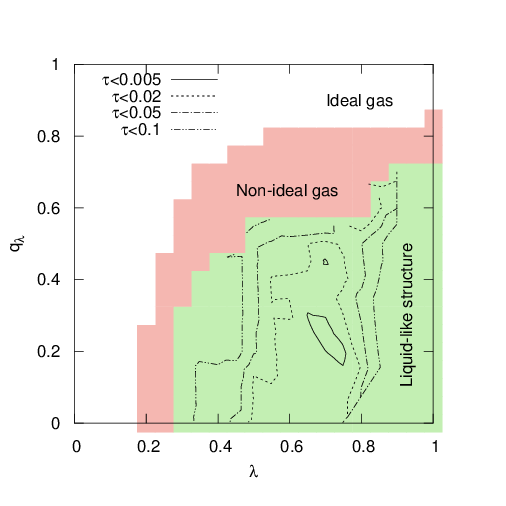}
\caption{Phase diagram of the core-shell-color string systems. Shaded regions indicate the structures that the system can adopt according to $(\lambda, q_\lambda)$: ideal gas (white region), non-ideal gas [pink (gray) shaded region], and liquid-like structure [green (light gray) shaded region]. The lines in the liquid-like region bound the pairs $(\lambda, q_\lambda)$ in which $\tau$ takes values less than  0.1 (dot-dot-dashed line), 0.05 (dot-dashed line), 0.02 (dotted line), and 0.005 (solid line).}
\label{fig:class}
\end{figure}

As we have a discrete set in the pairs ($\lambda$, $q_\lambda$), we interpolate $\tau$ with cubic splines to determine the regions wherein it takes values less than $\tau_0=$ $0.005, 0.02, 0.05, 0.1$. These results allow us to determine the regions where $T_c^*$ is bounded as
$(1-\tau_0)T_c<T_c^*<(1+\tau_0)T_c\,$.
Figure~\ref{fig:class} contains the counter lines of $\tau$ values discussed above, together with the obtained phase diagram for the pairs ($\lambda$, $q_\lambda$) according to: i) an ideal gas if the system shows the ideal gas structure for all inspected $\eta$, ii) a non-ideal gas if the system only shows a transition from the ideal gas to the non-ideal gas case, and iii) a liquid-like structure, if the system presents both ideal gas to non-ideal and non-ideal to liquid-like transitions.  

\begin{figure}[ht]
\centering
\includegraphics[scale=1]{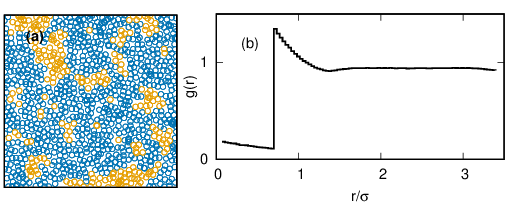}
\caption{(a) Example of a system that simultaneously exhibits a liquid-like structure and the spanning cluster has formed [blue (gray) disks] and (b) its corresponding $g(r)$. This corresponds to the pair $(\lambda$=0.7, $q_\lambda$=0.25) inside the region where $\tau<$0.005, with a filling factor near to the percolation threshold ($\eta$=0.861).}
\label{fig:perc-gr}
\end{figure}

Let us make contact with the experimental data. The critical temperature at which the QGP is formed takes values between 150 and 170 MeV for zero chemical potential. In this case, for  $\tau<0.005$  we have $|T_c-T_c^*|<1$ MeV, which means that both temperatures are very close. Then, our model predicts the liquid-like structure of the QGP (see  Fig.~\ref{fig:class}), which is in agreement with the experimental observations that suggest the liquid behavior of the QGP. In Fig.~\ref{fig:perc-gr} we show an example of the mentioned systems.

\section{Conclusions}\label{sec:conclusions}

In summary, we have presented a percolation-based string model that incorporates a repulsive interaction.
In this model, the systems can have three structures: ideal gas, non-ideal gas, and liquid-like. Our main result is the existence of systems that simultaneously show a gas-liquid transition and the emergence of a spanning cluster. Then, our model describes the experimentally observed liquid behavior of the QGP. We must remark that the inclusion of only the core region to generate hard-core systems ($q_\lambda$=0) is not enough to find the aforementioned systems. In our study, we used the color string percolation model, but the main results could be obtained in most of the string models if a repulsive interaction is incorporated properly.

In the simulations, we observed conglomerations of hard or soft strings that act as ``droplets" and ``bubbles," respectively. This is a consequence of a jamming-like effect produced by the hard strings. Then, there exists a particular filling factor from which can only be added  soft strings in the system. This could be an explanation of the second rise of $\varepsilon/T^4$ at  $T\sim$1.3-1.4 $T_c$ reported in Ref. \cite{PajaresArx}. 
In these kinds of systems, there would be an excess of soft strings at high $\eta$ values, which could indicate a new transition from liquid to gas and subsequently to an ideal gas. In the CSPM, at this temperature, the mean distance $d$ between the overlapping strings is  smaller than the string radius (computed like in Ref.~\cite{Castorina} for 2D systems, $d/r_0\sim$0.8, 0.9). For models that consider strings with cores, like our model, this means that the color field of the strings penetrates the core region until $\lambda\sim$0.4, 0.45, starting to see undressed color sources.

Finally, several questions remain open, for instance:

\textbullet  Notice that we worked with $\tau$, instead of calculating $T_c$ and $T_c^*$. For the temperatures, it is mandatory to determine $F(\eta)$ and $\langle p_T^2 \rangle_1$. With this information, it is possible to derive all of the phenomenology associated with the CSCSPM as a function of ($\lambda, q_\lambda$).
Even when we do not expect large changes, it would be interesting if the results shifted toward the experimental data.

\textbullet Since the third harmonic of the azimuthal distribution, $v_3$,  is very sensitive to the fluctuations of the string locations, we expect significant effects in the implementation of this model. In particular, as the relative weight of the edge is larger in small systems like pp or pA collisions than in heavy-ions collisions, the effects on $v_3$ can be larger.

\textbullet Our model opens the possibility of incorporating other interactions between strings. For example, it is well known that the strings can interact through a Yukawa-type potential, where the correlation length can be associated with the saturation scale $R_s=1/Q_s$ in the color glass condensate context. 

\textbullet To simulate a more realistic parton distribution we can implement different profiles in the CSCSPM, such as the Gaussian or Woods-Saxon distributions.

\begin{acknowledgments}
J.E.R. acknowledges financial support from Consejo Nacional de Ciencia y Tecnología (postdoctoral fellowship Grant No. 289198). B.D. was partially supported by a DGAPA-UNAM postdoctoral fellowship and acknowledges support from the CONEX-Plus programme funded by Universidad Carlos III de Madrid and the European Union's Horizon 2020 programme under the Marie Sklodowska-Curie grant agreement No. 801538. C.P. has received financial support from Xunta de Galicia (Centro singular de investigación de Galicia accreditation 2019-2022), by European Union ERDF, and by  the “María de Maeztu”  Units  of  Excellence program  MDM-2016-0692 and the Spanish Research State Agency.
We thank Nestor Armesto and David Vergara for their valuable comments.
\end{acknowledgments}

\bibliography{main}

\end{document}